\documentstyle[12pt]{article}
\newcommand{\be}{\begin{equation}}
\newcommand{\ee}{\end{equation}}
\newcommand{\bea}{\begin{eqnarray}}
\newcommand{\eea}{\end{eqnarray}}
\newcommand{\ba}{\begin{array}}
\newcommand{\ea}{\end{array}}
\newcommand{\bmat}{\left(\ba}
\newcommand{\emat}{\ea\right)}

\newcommand{\pole}{\frac{1}{\varepsilon}\frac{g^2}{16\pi^2}}

\begin{document}
\setlength{\baselineskip}{1pc}

\hbadness=10000
\begin{flushleft}
\rightline{KUCP-89}
\rightline{hep-ph/9603375}
\rightline{March, 1996}

\vspace*{0.5pc}
{\large\bf \hspace{1pc}Higher-twist Effects in Spin Structure Functions}
\footnote{Talk given at the RIKEN symposium on \lq Spin Structure of 
the Nucleon\rq,Wako, Japan, December 18-19, 1995}
\vspace{1pc}

{ \hspace{3pc}Tsuneo UEMATSU \\ \hspace{3pc}Department of Fundamental 
Sciences \\
 \hspace{3pc}FIHS, Kyoto University, Kyoto 606-01 \\ \hspace{3pc}JAPAN}
\end{flushleft}
%\vspace{0.5pc}
\leftline{\bf  Abstract}
\vspace{0.5pc}

 We discuss the QCD effects of the higher-twist operators
in the nucleon spin-dependent structure functions measured
by the polarized deep inelastic leptoproduction.
 We particularly study the renormalization of the twist-3 and
twist-4 operators within the framework of operator product 
expansions and renormalization group methods in perturbative QCD.
 Emphasis will be placed on the role of the operators proportional 
to equation of motion which appear in the mixing of the higher-twist 
operators through renormalization. 
 The logarithmic and power corrections in $Q^2$ due to the lowest spin 
higher-twist operators are discussed for the first moment of $g_1$ 
structure function, Bjorken and Ellis-Jaffe sum rules, as well as
the lower moments of $g_2$ structure function.

\vspace{1pc}
\leftline{\bf 1. Introduction}
\vspace{1pc}

In the last several years there has been a great deal of interest in nucleon's
spin structure functions $g_1(x,Q^2)$ and $g_2(x,Q^2)$ which can be measured
by deep inelastic scattering of polarized leptons on polarized nucleon targets.
From these polarized structure functions, we obtain the information on the
spin structure of quarks and gluons inside the nucleon, the dynamics of which
can be described by QCD. The information on spin content of nucleon can be 
also obtained from the polarized hadron-hadron collisions, such as direct
photon production, Drell-Yan process and so on.

So far the perturbative QCD has been tested for the effects of the leading 
twist operators, namely twist-2 operators for the unpolarized nucleon 
structure functions. QCD has been successful for describing the parton picture
of quarks and gluons corresponding to twist-2 effects in deep inelastic 
processes. On the other hand, there has been very little information 
about the higher-twist effects from the high energy deep inelastic processes.

Now, the spin structure functions provide us with a good place to study the 
higher-twist effects in the sense that i) The twist-3 operators contribute
to $g_2(x,Q^2)$ in the leading order of the scaling limit. ii)The twist-4
operator contributes to the moments of $g_1(x,Q^2)$, especially the first
moment relevant for the Bjorken and Ellis-Jaffe sum rules, in the order of 
$1/Q^2$. There we have $\overline{\psi}G\psi$ type twist-4 operator and no
four-fermi $\overline{\psi}\Gamma\psi\overline{\psi}\Gamma\psi$ type operator.
So the twist-4 effects in the polarized structure function reflect the
quark-gluon correlation inside the nucleon. 

\vspace{1pc}
\leftline{\bf 2. Twist-4 operator and the first moment of $g_1(x,Q^2)$}
\vspace{1pc}

In the framework of the operator product expansion and the renormalization
group method, we can derive sum rules for the first moment of the 
structure functions $g_1^{p,n}(x,Q^2)$, the Bjorken sum rule for the flavor
non-singlet combination, and the Ellis-Jaffe sum rule for the flavor singlet
component.  The twist-4 operator contributes to the
first moment of $g_1(x,Q^2)$ in the order of $1/Q^2$, and the coefficient
is determined by the nucleon matrix element of twist-4 operator. 
The logarithmic QCD correction to the twist-4 effects of order $1/Q^2$
is controlled by the anomalous dimension of the twist-4 operators. 

Now we investigate the renormalization of twist-4 operators
and calculate their anomalous dimensions which generate logarithmic 
$Q^2$ dependence. The result turns out to be as follows:
%\vspace{-0.3cm}
\bea
&&\hspace{-0.8cm}\Gamma_1^{p,n}(Q^2)\equiv \int_0^1 g_1^{p,n}(x,Q^2)dx 
\nonumber\\
&&\hspace{-0.8cm}=(\pm\frac{1}{12} g_A+\frac{1}{36}a_8)
\bigl ( 1-\frac{\alpha_s}{\pi}+ {\cal O}(\alpha_s^2)\bigr )
+\frac{1}{9}\Delta\Sigma\bigl ( 
1-\frac{33-8n_f}{33-2n_f}\frac{\alpha_s}{\pi}+{\cal O}(\alpha_s^2)
\bigr )  \nonumber\\     
&&\hspace{-0.8cm}-\frac{8}{9Q^2}\Bigl [ \{ \pm\frac{1}{12} f_3+
\frac{1}{36}f_8\}
\left(\frac{\alpha_s(Q_0^2)}{\alpha_s(Q^2)}\right)^
{-\frac{\gamma_{NS}^0}{2\beta_0}}\hspace{-0.1cm}+\frac{1}{9}f_0
\left(\frac{\alpha_s(Q_0^2)}{\alpha_s(Q^2)}\right)^
{-\frac{1}{2\beta_0}
(\gamma_{NS}^0+\frac{4}{3}n_f)}\Bigr ]
\eea
%\vspace{-0.3cm}
where $g_1^{p(n)}(x,Q^2)$ is the spin structure function of the 
proton (neutron) and the plus (minus) sign is for proton (neutron), with 
$x$ and $Q^2$ being the Bjorken 
variable and the virtual photon mass squared. On the right-hand side, 
$g_A \equiv G_A/G_V$ is the ratio of the axial-vector to vector coupling 
constants. 
Here we assume the number of active flavors in the current region of
$Q^2$ is $n_f=3$. 
$a_8$ and $a_0=\Delta\Sigma$ are flavor-$SU(3)$ octet and singlet
parts, as given by
$a_8 \equiv \Delta u +\Delta d -2\Delta s$, 
$\Delta\Sigma \equiv \Delta u +\Delta d +\Delta s$.
Here we have suppressed the target mass effects, which can
be taken into account by the Nachtmann moments \cite{HKU}.
The only relevant twist-4 operator $R_{2\sigma}^i$ is of the form bilinear 
in quark fields and linear in the gluon field strength, and is given by, 
together with the nucleon matrix element as
\be
R_{2\sigma}^i=g\overline{\psi}\tilde{G}_{\sigma\nu}\gamma^{\nu}t^i\psi,
\quad \langle p,s|R_{2\sigma}^i|p,s \rangle=f_i s_{\sigma} 
\ee
where $\tilde{G}_{\mu\nu}=\frac 1 2 \varepsilon_{\mu\nu\alpha\beta}
G^{\alpha\beta}$ is the gluon dual field strength, $t^i$ is the flavor 
matrix and $s_\mu$ is the nucleon covariant spin vector. 
$f_0$, $f_3$ and $f_8$ are the twist-4 counter parts of $a_0$, $a_3$ and
$a_8$. $f_i$'s are scale dependent and here they are those at $Q_0^2$.

The common feature for the renormalization of higher-twist operators
is that there appear a class of operators proportional to equations of 
motion, which we call EOM operators. And there exists the operator mixing 
among twist-4 operators including EOM operators through renormalization. 
The composite operators are renormalized as $(O_i)_R=\sum_j Z_{ij}(O_j)_B$.
There are five possible spin-1, twist-4 operators are as follows \cite{KUKY} 
\bea
 && R_1^\sigma = -\overline{\psi}\gamma_5
       \gamma^{\sigma}D^2\psi  , \quad                 
  R_2^\sigma = g \overline{\psi}
       \tilde{G}^{\sigma\mu}\gamma_{\mu}\psi            \nonumber\\
 && E_1^\sigma = \overline{\psi} \gamma_5
          \not{\!\!D} \gamma^{\sigma} \not{\!\!D} \psi 
         -\overline{\psi} \gamma_5 D_\sigma\not{\!\!D}\psi
         -\overline{\psi} \gamma_5 \not{\!\!D} D^\sigma\psi  \\
 && E_2^\sigma = \overline{\psi} \gamma_5\partial^{\sigma}
          \not{\!\!D}\psi + \overline{\psi} \gamma_5 \not{\!\!D}
          \partial^{\sigma}\psi, \quad                          
  E_3^\sigma = \overline{\psi} \gamma_5 \gamma^\sigma \not{\!\partial}
          \not{\!\!D}\psi + \overline{\psi} \gamma_5
          \not{\!\!D}\not{\!\partial}\gamma^\sigma\psi
            \nonumber 
\eea
where $D_\mu=\partial_\mu-igA_\mu^aT^a$ is the covariant derivative.
Using the identities, 
$D_\mu=\frac{1}{2}\{\gamma_\mu,\not{\!\!D}\}$ and
$[ D_\mu ,D_\nu ]=-igG_{\mu\nu}$, we obtain the constraint
$R_1^{\sigma}=R_2^{\sigma}+E_1^{\sigma}$.
If we take a basis of independent operators as ($R_2$, $E_1$, $E_2$, $E_3$),
we have the following renormalization matrix
\be
\bmat{c} 
R_2 \\ E_1 \\ E_2 \\ E_3 
\emat_{\hspace{-0.1cm}R}
=
\bmat{cccc}
Z_{11} & Z_{12} & Z_{13} & Z_{14} \\
 0     & Z_{22} & Z_{23} & Z_{24} \\
0 & 0 & Z_{33} & 0 \\
0 & 0 & 0 & Z_{44} 
\emat
\bmat{c} 
R_2 \\ E_1 \\ E_2 \\ E_3 
\emat_{\hspace{-0.1cm}B}
\label{n3z1}
\ee
where $R$($B$) denotes the renormalized (bare) quantities.
The structure of the renormalization matrix is consistent with the general
theory; (i)The counter terms for the EOM operators are given by the the EOM 
operators themselves. (ii) A certain type of operators do not get renormalized.
(iii)The gauge variant operators also contribute to the mixing.
Since a physical matrix element of EOM operators vanishes\cite{POLI}, 
the only operator 
which really contribute to the physical matrix element is $R_2$.	
This twist-4 operator corresponds to the trace part of twist-3 operator,
$(R_{\tau=3})_{\sigma\mu_1\mu_2}=g\overline{\psi}\tilde{G}_{\sigma\{\mu_1}
\gamma_{\mu_2\}}\psi - $traces.
We compute $Z_{ij}$ by evaluating the off-shell Green's function
of twist-4 composite operators keeping the EOM operators as independent 
operators.
Thus we can avoid the subtle infrared divergence which may appear in the
on-shell amplitude with massless particle in the external lines.
${\left(\Gamma_{O_i}\right)}_R =\sum_jZ_2\sqrt{Z_3}Z_{ij}
\left (\Gamma_{O_j}\right )_B$, 
where $Z_2$ and $Z_3$ are wave function renormalization constants for
quarks and gluon fields. Writing $Z_{ij}\equiv \delta_{ij} +
{1\over\varepsilon}{{g^2}\over{16\pi^2}}z_{ij}$, we obtain $z_{ij}$
\cite{KUKY}.
The result is in agreement with the general theorem on the renormalization
mixing matrix \cite{COLL}; i.e. the mixing matrix is triangular.
And the couter terms for the EOM operators are those
from the EOM themselves. $E_2, E_3$ are confimed to be free from 
renormalization as implied by the general theorem.
We also note that gauge variant EOM operator is necessary for the 
renormalization.

Therefore  the anomalous dimension $\gamma_{R_2}$ turns out to be
( $C_2(R)=4/3$ )
\be
\gamma_{R_2}(g)=\frac{g^2}{16\pi^2}\cdot 2z_{11}+O(g^4), \quad 
\gamma_{NS}^0=2z_{11}=\frac{16}{3}C_2(R)
\ee
which coincides with the result obtained by Shuryak and Vainshtein \cite{SV}
in a different method.

Let us now turn to the flavor singlet component. The possible non vanishing 
twist 4 and spin 1 gluon operators is the following using the equation of 
motion: 
\be
\tilde{G}^{\alpha\sigma}{D}^{\mu}G_{\mu\alpha}=g\overline{\psi}
\gamma_{\alpha}\tilde{G}^{\alpha\sigma}\psi
\ee
So we have to take into account the mixing between $R_2^{\sigma}
=g\overline{\psi}\gamma_{\alpha}\tilde{G}^{\alpha\sigma}\psi$
and
\be
E_G^{\sigma}=\tilde{G}^{\alpha\sigma}D^{\mu}G_{\mu\alpha}-
g\overline{\psi}\gamma_{\alpha}\tilde{G}^{\alpha\sigma}\psi
\ee
The mixing matrix element between $R_2$ and $E_G$ 
turns out to be $Z_{15}=\pole\times\frac 2 3 n_f$.
And we get $Z_{11}^{S}=Z_{11}^{NS}+\frac 2 3 n_f$, i.e.
$\gamma_S^0=\gamma_{NS}^0+\frac 4 3 n_f$. 

\vspace{1pc}
\leftline{\bf 3. Twist-3 operators and $g_2(x,Q^2)$}
\vspace{1pc}

Let us now consider the operator mixing of twist-3 operators for $n=3$ case
including EOM operators as follows,
\bea
&& R_F^{\sigma\mu_{1}\mu_{2}} =i^2S'\overline{\psi}\gamma_5
       \gamma^{\sigma}D^{\mu_1}D^{\mu_{2}}\psi-\mbox{traces},\quad
  R_G^{\sigma\mu_{1}\mu_{2}}=g\overline{\psi}\tilde{G}^{\sigma\{\mu_1}
\gamma^{\mu_2\}}\psi-\mbox{traces} \nonumber\\
&&  R_m^{\sigma\mu_{1}\mu_{2}} =im\overline{\psi}\gamma_5\gamma^{\sigma}
      D^{\{\mu_1}\gamma^{\mu_2\}}\psi-\mbox{traces}, \nonumber\\
&&R_{eq}^{\sigma\mu_{1}\mu_{2}} =
           i \frac{1}{3} S' [ \overline{\psi} \gamma_5
          \gamma^{\sigma} D^{\mu_1}\gamma ^{\mu_{2}} 
(i\not{\!\!D} - m )\psi + \overline{\psi} (i\not{\!\!D} - m )
              \gamma_5 \gamma^{\sigma} D^{\mu_1} \gamma ^{\mu_{2}} \psi ] 
\nonumber 
\eea
where $S'$ means the anti-symmetrization between $\mu_i$ and $\sigma$,
and symmetrization in $\mu_1$ and $\mu_2$.
Now we notice~\cite{SVETAL,JAFFE} that these operators are
related through  EOM operators; $R_F=\frac 2 3 R_m+R_G+R_{eq}$.
If we eliminate $R_F$, we solve the mixing among $R_G$, $R_m$, $R_{eq}$ and
$R_{eq1}$, which is a gauge-variant operator obtained by replacing $D^{\mu_1}$
with $\partial^{\mu_1}$. 
The $n=3$ moment of $g_2(x,Q^2)$ is given by
\be
M_3(Q^2)\equiv \int_0^1dx x^2 g_2(x,Q^2)= -\frac 1 3 a_3 E_q^3(Q^2)+
\frac 1 2 d_3 E_G^3(Q^2) + \frac 1 2 e_3 E_m^3(Q^2)
\ee
where the nucleon matrix elements of the independent operators, $R_q^{n=3}$ 
(twist-2 operator), $R_G^{n=3}$ and $R_m^{n=3}$ are denoted by $a_3$, $d_3$
and $e_3$, respectively. The coefficient functions in this basis have the tree
values $E_q^3(\mbox{tree})=1$, $E_G^3(\mbox{tree})=1$ and 
$E_m^3(\mbox{tree})=2/3$ and their evolution in $Q^2$ is determined by the 
anomalous dimensions obtained by the renormalization matrix. 
The mixing problem has been studied keeping the EOM operators and evaluating 
the off-shell Green's functions for general spin $n$ case \cite{KTUY}. The
result should be compared with that obtained by Ji and Chou \cite{JC} based
on a different method.

\vspace{1pc}
\leftline{\bf 4. Concluding remarks}
\vspace{1pc}
Although, in this talk, we have confined ourselves to spin structure 
function measured by polarized leptoproduction, the information on nucleon
spin structure can as well be obtained by polarized hadron-hadron collisions 
like direct photon production, Drell-Yan process etc.
There has been a QCD analysis of chiral-odd twist-3 structure function 
$h_L(x,Q^2)$ in the polarized Drell-Yan process \cite{KT}. By computing
the off-shell Green's functions with a suitable projection, the authors have 
solved the operator mixing problem for the general spin.   
Here we also note that the single spin asymmetry observed in the direct
photon process, $p^{\uparrow}+p\rightarrow \gamma +X$ can be related to
the second moment of ${\bar g}_2$, which is the twist-3 part of $g_2$.

%Here we would like to emphasize that both $g_1$ and $g_2$ are needed to
%study the target mass effects \cite{HKU} in order to isolate the higher-twist 
%effects in the moment sum rules of structure functions based on QCD.
Note that the higher-twist contribution suffers from ambiguity due to
renormalon singularity \cite{JEMK}, while its logarithmic $Q^2$ dependence 
is free from such ambiguity.
The nucleon matrix elements of twist-4 operators can be in principle, 
separated by extracting $Q^2$ dependence and the target dependence which
is seen by eq.(1). This should be carried out with the more accurate data
in the future experiments at CERN, SLAC, HERMES at DESY and RHIC.

\vspace{1cm}
I would like to thank H. Kawamura, J. Kodaira, K. Tanaka and Y. Yasui
for valuable discussions.
%%%%%%%%%%%%%%%%%%%%%%%%%%%%%%%%%%%%%%%%%%%%%%

\end{document}